\documentclass[pre,twocolumn,showpacs,amsmath,amssymb]{revtex4}
\usepackage{mathrsfs}
\usepackage{graphicx} % Include figure files
\usepackage{dcolumn}  % Align table columns on decimal point
\usepackage{bm}       % bold math

\newcommand{\bsub}{\begin{subequations}}
\newcommand{\esub}{\end{subequations}}
\newcommand{\rrr}{\mathbf{r}}
\newcommand{\iii}{\mathbf{i}}

\newcommand{\vvv}{\mathbf{v^{{}}}}
\newcommand{\FFF}{\mathbf{F^{{}}}}
\newcommand{\lamD}{\lambda^{{}}_D}
\newcommand{\lamDsqr}{\lambda^{{2}}_D}
\newcommand{\oD}{\omega^{{}}_D}
\newcommand{\oC}{\omega^{{}}_c}
\newcommand{\nnp}{n^{+_{}}}
\newcommand{\nnm}{n^{-_{}}}
\newcommand{\npm}{n^{\pm_{}}}

\newcommand{\pp}{\partial^{{}}}
\newcommand{\nablabf}{\boldsymbol{\nabla}}
\newcommand{\kB}{k^{{}}_\textrm{B}}

\newcommand{\Vext}{V^{{}}_\textrm{ext}}

\newcommand{\ie}{\textit{i.e.}}
\newcommand{\eg}{\textit{e.g.}}

\begin{document}

\title{Electro-hydrodynamics of binary electrolytes driven by modulated surface
potentials}

\author{Niels Asger Mortensen,$^1$ Laurits H{\o}jgaard Olesen,$^1$
Lionel Belmon,$^{1,2}$ and Henrik Bruus$^1$}

\affiliation{$^1$MIC -- Department of Micro and Nanotechnology,\\
Technical University of Denmark, DK-2800 Kongens Lyngby, Denmark\\
$^2$Ecole Centrale de Nantes, F-44321 Nantes, France }

\date{March 8, 2005}
\begin{abstract}
We study the electro-hydrodynamics of the Debye screening layer
that arises in an aqueous binary solution near a planar insulating
wall when applying a spatially modulated AC-voltage. Combining
this with first order perturbation theory we establish the
governing equations for the full non-equilibrium problem and
obtain analytic solutions in the bulk for the pressure and
velocity fields of the electrolyte and for the electric potential.
We find good agreement between the numerics of the full problem
and the analytics of the linear theory. Our work provides the
theoretical foundations of circuit models discussed in the
literature. The non-equilibrium approach also reveals unexpected
high-frequency dynamics not predicted by circuit models.
\end{abstract}

\pacs{47.65.+a, 47.32.-y, 47.70.-n, 85.90.+h} \maketitle

%%%%%%%%%%%%%%%%%%%%%%%%%%%%%%%%%%%%
\section{Introduction}
\label{sec:introduction}

Recently, there has been quite some interest in
electro-hydrodynamics in microfluidic systems. AC-driven,
modulated surface potentials have been used for pumping, fluid
circulation, and
mixing~\cite{yeh:97a,Ajdari:00a,Brown:00a,Green:00a,Gonzalez:00a,Green:02a,Nadal:02a,Ajdari:02a,Gitlin:03a,Studer:04b,Cahill:04a,Arnoldus:04a,Hansen:04a}.
For an overview of AC electro-osmosis we refer to
Refs.~\onlinecite{Ramos:98a,Morgan:03a,Stone:04a,Bazant:04b,Squires:04a,Bazant:04a}~and
references therein.

\begin{figure}[b!]
\centerline{\includegraphics[width=\columnwidth]{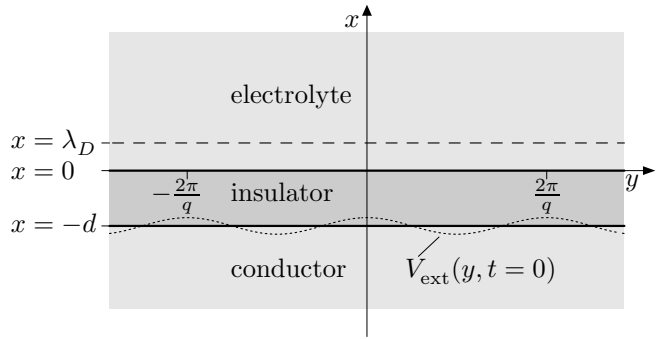}}
\caption{A sketch of the system under study. The binary
electrolyte is situated in the half space $x>0$. Below it, for
$-d<x<0$, is a planar wall consisting of an insulating dielectric
slab of thickness $d$ and below that, for $x<-d$, is a
semi-infinite conductor. The top surface, $x=-d$, of the conductor
is biased by a periodically modulated potential $\Vext(y,t)$ of
period $2\pi/q$ (dotted line), which gives rise to the formation
of a Debye screening layer of thickness $\lamD$ in the electrolyte
(dashed line).} \label{fig:setup}
\end{figure}

Although AC electro-osmosis is typically analyzed with the help of
homogeneous circuit elements open questions remain about the
applicability of such approaches \cite{Bazant:04a}. We revisit the
problem studied by Ajdari~\cite{Ajdari:00a} where an electrolyte
is perturbed by an AC-driven spatially modulated surface
potential, but include explicitly an insulating layer between the
electrode providing the driving potential and the electrolyte. We
primarily think of this insulator as an oxide grown intentionally
for device purposes, but it could also represent the molecular
Stern layer in case of non-oxidized electrodes. We develop a full
non-equilibrium description of the electro-hydrodynamics of this
system thus extending previous modeling of the surface and the
Debye layer as simple capacitors. This allows us to study the full
dynamics of ion concentrations, electrical potentials, velocity
fields, pressure gradient-fields, and electrical currents as well
as the justifications for a description based on homogeneous
circuit elements.

In the following we consider a binary electrolyte, \ie, an aqueous
solution of a salt containing a positive and a negative type of
ions with charges $+Ze$ and $-Ze$, respectively, where $Z$ is the
valence and $e$ the elementary charge. In terms of Cartesian
coordinates $xyz$ the electrolyte is confined to the semi-infinite
space $x>0$ by an impenetrable, homogeneous and planar insulating
layer with dielectric constant $\epsilon_s$ placed at $-d<x<0$,
see Fig.~\ref{fig:setup}. The insulating layer is bounded by a
conductor at $x<-d$ which has been biased at the surface $x=-d$ by
a spatially modulated, external AC potential $\Vext(y,t)$
\begin{equation}
\label{eq:Vext}
\Vext(y,t)=V_0\cos(qy)e^{i\omega t},
\end{equation}
where $V_0$ is the amplitude, $q$ the wavenumber of the spatial
modulation, and $\omega$ the driving angular frequency.

There is complete translation invariance along the $z$ axis, so
the $z$ coordinate drops out of our analysis, and all positions
$\rrr = x \textbf{e}_x + y\textbf{e}_y$ are therefore just
referring to the $xy$ plane.

The manuscript is organized as follows: in
Sec.~\ref{sec:Non-equilibrium description} we present the
non-equilibrium description and in Secs.~\ref{sec:statics} and
\ref{sec:dynamics} we analytically study linearized equations of
the static and dynamic regimes, respectively. In
Sec.~\ref{sec:FullNumerics} we study numerical solutions of the
fully coupled non-linear electro-hydrodynamic problem. Finally, in
Sec.~\ref{sec:discussion} we compare these solutions with the
analytical solutions of the linearized equations, and furthermore
contrast our results with the literature, before we in
Sec.~\ref{sec:conclusion} conclude.

\section{Non-equilibrium description}
\label{sec:Non-equilibrium description}

The basic non-equilibrium formalism for continuum
electro-hydrodynamics is well-known (see, \eg,
Ref.~\onlinecite{Bazant:04a}), but as mentioned in the
introduction we explicitly include an insulating layer in the
description. We do not include an intrinsic zeta-potential, \ie,
no un-passivated surface charges on the insulator-electrolyte
interface. We note that experimentally any intrinsic homogeneous
zeta-potential may be compensated by a corresponding DC-shift
added to the applied AC potential. Due to heterogeneous surfaces,
it may be anticipated that although the average zeta-potential is
nulled out, there might be fluctuations left. These will be the
topic for future work. For zero intrinsic zeta-potential we solve
the full non-linear equations numerically, but to obtain
analytical result we also study the linearized equation with
special emphasis on the capacitance due to the insulating layer.
In Sec.~\ref{sec:FullNumerics} we show that for experimental
relevant parameters, the linear theory is surprisingly good.
Typical values of various central parameters are listed in
Table~\ref{tab:parameters}.

\subsection{The insulating layer, $\boldsymbol{-d<x<0}$}

The insulating layer contains neither free space charge nor free
currents so the electrical potential $\phi(\rrr,t)$ is governed by
the Laplace equation,
\begin{equation}\label{eq:Laplace}
\nablabf^2\phi(\rrr,t) = 0, \quad \textrm{for}\; -d<x<0.
\end{equation}

\subsection{The electrolyte, $\boldsymbol{x>0}$}

In the liquid electrolyte we consider the ionic densities
$\npm(\rrr,t)$, the potential $\phi(\rrr,t)$, the ionic current
densities (the ionic flux densities) $\iii^\pm(\rrr,t)$, the
velocity field $\vvv(\rrr,t)$ of the electrolyte, and the pressure
$p(\rrr,t)$. In the following we suppress $(\rrr,t)$ unless needed
for clarity.

The number densities of the ions couple to the potential via
Poisson's equation,
\begin{subequations}
\begin{equation} \label{eq:Poisson}
\nablabf^2 \phi = - \frac{Ze}{\epsilon}\big(\nnp - \nnm\big).
\end{equation}

The ionic current densities are coupled to the ionic densities by
a continuity equation, which in the absence of any chemical
reactions in the system is
\begin{equation} \label{eq:conteq}
\pp_t \npm = -\nablabf \cdot \iii^\pm.
\end{equation}

The presence of convection or of gradients in the densities $\npm$
and the electric potential $\phi$ will generate ionic current
densities $\textbf{i}^{\pm}$. The Nernst--Planck equation gives
these currents
\begin{equation} \label{eq:NP}
\iii^\pm =  - D \nablabf \npm  + \npm\vvv  \mp \mu \npm \nablabf
\phi,
\end{equation}
where, for simplicity, we have assumed that the two types of ions
have the same diffusivity $D$ and the same mobility $\mu$. We
remind the reader that both the diffusivity $D$ and the electric
conductivity $\sigma$ are linked to the mobility $\mu$ via the
Einstein relation $D =(\kB T/Ze) \mu$ and $\sigma^{\pm_{}} = Z e
n^\pm \mu$.

Finally, the velocity field and pressure of the liquid are coupled
to the potential and ionic densities by the Navier--Stokes
equation
 \label{Navier-Stokes_1st}
 \begin{equation} \label{eq:NS}
 \rho \big[\pp_t \vvv + (\vvv \cdot \nablabf) \vvv \big]
 = -\nablabf p + \eta \nablabf^2 \vvv
 - Z e\big[\nnp - \nnm\big] \nablabf \phi,
 \end{equation}
%
%
%\label{Navier-Stokes_1st}
%\begin{align}
%&\rho \big[\pp_t \vvv + \vvv \cdot \nablabf
%\vvv \big] =-\nablabf p\label{eq:NS} \\
%&\quad+ \eta \nablabf^2 \vvv - Z e\big[\nnp - \nnm\big] \nablabf
%\phi,\nonumber
%\end{align}
%
where $\rho$ is the mass density, $\eta$ is the viscosity of the
liquid, and $p$ is the pressure. Furthermore, treating the
electrolyte as an incompressible fluid we have
\begin{equation}\label{eq:incompressible}
\nablabf \cdot \vvv=0.
\end{equation}
\label{eq:basic}
\end{subequations}
The coupled field-equations, Eqs.~(\ref{eq:Poisson})
to~(\ref{eq:incompressible}), fully govern the physical fields
$n^{\pm}$, $\phi$, $\textbf{i}^\pm$, $\bf v$, and $p$.

\begin{table}[t!]
\caption{Typical values of central parameters.}
\label{tab:parameters}
\begin{tabular}{lcr@{}l}\hline
Spatial modulation          & $q^{-1}$                                  &  $10^{-5}$&~m \rule{0mm}{4mm} \\
Insulator thickness         & $d$                                       &  $10^{-8}$&~m \\
Debye length                & $\lambda_D$                               &  $10^{-8}$&~m  \\[1mm]
\hline
Resonance frequency         & $\omega^*$                                &  $10^5$&~rad/s \rule{0mm}{4mm}\\
Debye frequency             &$\oD=\sigma_{\scriptscriptstyle \infty}/\epsilon$                    &  $10^7$&~rad/s \\
Critical frequency          &$\oC=(\eta/\rho)\: q^2$                  &  $10^{4}$&~rad/s\\[2mm]
\hline
Thermal voltage             &$V_T=(1+\delta)\kB T/Ze$                              &  250&~mV\rule{0mm}{4mm}\\[1mm]
Convective voltage            & $V_c=\sqrt{(1+\delta)\eta D/\epsilon}$    &  100&~mV \\[2mm]
\hline
Ionic density               & $n_{\scriptscriptstyle \infty}^{{}}$      &  1&~mol\,L$^{-1}$ \rule{0mm}{4mm}\\
Viscosity                   &$\eta$                                     &  ${10}^{-3}$&~Pa\,s \\
Mass density                &$\rho$                                     &  $10^3$&~kg\,m$^{-3}$\\
Ionic diffusivity           &$D$                                        &  $10^{-9}$&~m$^{2}$\,s$^{-1}$ \\
Capacitance ratio           &$\delta=C_D/C_s$                           &  10& \\[1mm]
\hline
\end{tabular}
\end{table}

\subsection{Boundary conditions}

Assuming a vanishing zeta-potential the boundary condition for the
electric potential is
\bsub
\begin{align}
\label{eq:BCphiSurf}
\phi(\rrr,t)\big|_{x=-d} & = V_{\rm ext}(y,t),\\
\label{eq:BCphiInfty} \phi(\rrr,t)\big|_{x=\infty} &= 0.
\end{align}
\esub
At the interface between the
electrolyte and the insulating region the normal component of the
ionic current density vanishes,
\begin{equation}\label{boundary_current}
0=\partial_x \npm(\rrr,t)\big|_{x=0} \pm \frac{Ze}{\kB T}
\npm(\rrr,t)
\partial_x \phi(\rrr,t)\big|_{x=0}.
\end{equation}
Here, we have utilized Eq.~(\ref{eq:NP}) and the absence of
convection at the interface due to the no-slip boundary condition,
\begin{equation}\label{eq:noslip}
\vvv(\rrr,t)\big|_{x=0}=\textbf{0}.
\end{equation}
For the ionic densities we have
\begin{equation}\label{eq:boundary_density}
\npm(\rrr,t)\big|_{x=\infty} = n_{\scriptscriptstyle \infty}^{{}},
\end{equation}
where $n_{\scriptscriptstyle \infty}^{{}}$ is the homogeneous
density of either of the two types of ions in the absence of an
external perturbation, \ie, when $V_0 = 0$. For the pressure, we
assume that we have no externally applied pressure gradients so
that $p$ is the internal pressure caused by fluid flow and the
electrical forces on the ions.

\section{Static regime, $\boldsymbol \omega$ = 0}
\label{sec:statics}

In the static regime we have equilibrium and neither current nor
fluid flow, \ie,  $\textbf{i}^\pm=\textbf{0}$ and
$\vvv=\textbf{0}$. The pressure gradient balances the electrical
forces on the charges. The governing equations for $\phi$ and
$n^\pm$ of course reduce to those of electro-statics.

In the insulating layer it follows from Eqs.~(\ref{eq:Laplace})
and~(\ref{eq:BCphiSurf}) that
\begin{equation} \label{eq:phiStaticIns}
\phi(\rrr) = \Big[ \mathcal{B}^{{}}_1 e^{-qx} + \mathcal{B}^{{}}_2
e^{qx}\Big]\: \cos(qy), \quad \textrm{for}\; -d<x<0,
\end{equation}
where $\mathcal{B}^{{}}_{1}$ and $\mathcal{B}^{{}}_{2}$ are
integration constants.

In the electrolyte $\phi(\rrr)$ is governed by the nonlinear
Poisson--Boltzmann equation~\cite{Bazant:04a}
\begin{equation} \label{eq:PoissonBoltzmann}
\nablabf^2\phi(\rrr) = \frac{\kB
T}{Ze\lambda_D^{2}}\sinh\left[\frac{Ze}{\kB T}\:
\phi(\rrr)\right], \quad \textrm{for}\; x>0,
\end{equation}
introducing the Debye screening length
\begin{equation}
\lamD \equiv \sqrt{\frac{\epsilon \kB T}{2 Z^2 e^2
n_{\scriptscriptstyle \infty}^{{}}}}.
\end{equation}
For $q$ going to zero we have a constant surface potential
\begin{subequations}
\begin{equation}
\lim_{q\rightarrow 0}\phi(\rrr)\big|_{x=0} \equiv \phi_0
\end{equation}
and the solution to Eq.~(\ref{eq:PoissonBoltzmann}) is given by
the well-known Gouy--Chapman solution~\cite{Bazant:04a}
\begin{equation} \label{eq:GouyChapman}
\lim_{q\rightarrow 0} \phi^{{}}(\rrr) = \frac{4 \kB T}{Ze}\:
\textrm{arctanh} \Bigg[ \tanh\bigg(\frac{Ze\phi_0}{4\kB T}\bigg)\:
e^{-x/\lamD} \Bigg].
\end{equation}
\end{subequations}
For $q\neq 0$ we are not aware of any analytical solutions, but as
we shall show, analytical results can be obtained in the
Debye--H\"{u}ckel approximation $Ze\phi\ll \kB T$, where
Eq.~(\ref{eq:PoissonBoltzmann}) becomes linear,
\begin{equation}
\nablabf^2\phi(\rrr) = \lambda_D^{-2}\phi(\rrr), \quad
\textrm{for}\; x>0.
\end{equation}
Here, the corrections are to third order in $Ze\phi/\kB T$ because
$\sinh(x)=x+{\cal O}\big(x^3\big)$.
The space charge follows from Poisson's equation,
Eq.~(\ref{eq:Poisson}). From a straightforward solution for $\phi$
and $Ze(n^+-n^-)$ we arrive at the following expression relating
the total potential drop across the system and the accumulated
charge in the electrolyte,
\bsub
\begin{equation}\label{eq:Ceff}
\phi(\infty,y)-\phi(-d,y) \equiv C^{-1}_{\rm eff} \int_0^\infty dx
Ze[n^+(\rrr)-n^-(\rrr)].
\end{equation}
The coefficient,
 \begin{align}
 C_{\rm eff}^{-1}&= \big[1+(q\lambda_D)^2\big]\frac{\sinh(qd)}{qd}\:
 C_s^{-1}\nonumber\\
 &\quad
 +\sqrt{1+(q\lambda_D)^2}\:\cosh(qd)\:C_D^{-1},\label{eq:Ceff2}
 \end{align}
is identified as the inverse of an effective series capacitance.
The constant $C_s$ is the intrinsic surface capacitance and $C_D$
the capacitance of the Debye layer given by
\begin{align}
C_s&\equiv\frac{\epsilon_s}{d},\\
C_D&\equiv\frac{\epsilon}{\lambda_D}.
\end{align}
In Ref.~\onlinecite{Ajdari:00a} the potential in the bulk of the
electrolyte ($x\gg \lambda_D$) is governed by the Laplace
equation, which is coupled to the external potential $V_{\rm ext}$
by an effective capacitance $C_0$ given by
\begin{equation} \label{C0def}
C_0\equiv\left(C_s^{-1}+C_D^{-1}\right)^{-1}.
\end{equation}
\esub
It follows from Eq.~(\ref{eq:Ceff2}) that this approach for
$\omega=0$ is valid up to second order in the parameters
$q\lambda_D$ and $qd$.

\section{Linearized dynamic regime,
$\boldsymbol \omega \;\mathbf{>}$ 0}
\label{sec:dynamics}

We now solve Eqs.~(\ref{eq:basic}) in the dynamic regime,
$\omega>0$. First the ionic current densities are eliminated by
inserting Eq.~(\ref{eq:NP}) into Eq.~(\ref{eq:conteq}). Using the
incompressibility of the fluid, Eq.~(\ref{eq:incompressible}), we
get the continuity equation
 \begin{equation} \label{eq:conteq3}
 \pp_t \npm = D\: \nablabf^2 \npm -
 \big(\nablabf \npm\big)\cdot \vvv
 \pm \mu \nablabf \cdot \big(\npm\:
 \nablabf \phi \big).
 \end{equation}

\subsection{Debye--H\"{u}ckel approximation}
\label{Sec:debye-huckel}

To advance further by analytical methods, we now linearize the
continuity equation, Eq.~(\ref{eq:conteq3}), in the density as
follows. We assume $\npm(\rrr,t)\big|_{x=\infty} \equiv
n_{\scriptscriptstyle \infty}^{{}}$ and write
\begin{equation} \label{eq:n0var}
\npm(\rrr,t) = n_{\scriptscriptstyle \infty}^{{}} +
\delta\npm(\rrr,t), \qquad
\lim_{x\rightarrow\infty}\delta\npm(\rrr,t)=0.
\end{equation}
Since we assume a zero intrinsic zeta-potential it is a non-zero
$V_0$ that spawns $\delta\npm \neq 0$, and when the applied
voltage $V_0$ is much smaller than the thermal voltage $V_T$,
defined by $V_T \equiv (1+C_D/C_s)\kB T/Ze $ (as we shall see in
the next subsection), we have $|\delta\npm| \ll
n_{\scriptscriptstyle \infty}^{{}}$. In this limit the
Debye--H\"{u}ckel approximation is valid, and $\npm\: \nablabf
\phi$ is substituted by $n_{\scriptscriptstyle \infty}^{{}}\:
\nablabf \phi$ in Eq.~(\ref{eq:conteq3}). We subsequently use
Eq.~(\ref{eq:Poisson}) to replace $\nablabf^2\phi$ with
$-Ze\nu/\epsilon$ where
\begin{equation}
\nu\equiv n^+-n^-=\delta n^+-\delta n^-.
\end{equation}
Finally, we form the difference of the "$\pm$"-versions of
Eq.~(\ref{eq:conteq3}) and obtain the partial differential
equation
\begin{equation}\label{eq:densitydifference}
\pp_t \nu = \left[  D\nablabf^2 -D\frac{1}{\lamDsqr} -\vvv\cdot
\nablabf\right] \nu.
\end{equation}

\subsection{Diffusive regime}
\label{Sec:diffusion}

Our study of the static regime reconfirm the well-known result
that the net charge density is non-zero only in the Debye layer,
$x\lesssim 3\lambda_D$. In this region convection will be
suppressed due to the no-slip boundary condition. Thus, convection
can be neglected, diffusion will dominate (corresponding to a low
P\'{e}clet number), and the electrodynamics can be solved
independently of the hydrodynamics. On the other hand, the
hydrodynamics of course still depends on the electrodynamics via
the body-force. Since the density difference $\nu$ changes over
the length scales $\lambda_D$ and $q^{-1}$ for the $x$ and $y$
directions, respectively, the condition for the decoupling is
$|v_x|/\lambda_D+|v_y|q\ll D q^2$ for $0<x\lesssim3\lamD$. In this
limit Eq.~(\ref{eq:densitydifference}) has a general
$\cos(qy)e^{i\omega t}$ modulated decaying solution of the form
\bsub
\begin{equation}\label{eq:nu_general}
\nu= \mathcal{C}_1 e^{-\kappa x}\cos(qy)e^{i\omega t},\quad x>0,
\end{equation}
where the decay parameter $\kappa$ depends on the ratio between
the frequency $\omega$ and the Debye frequency $\oD$,
\begin{align}
\kappa &\equiv \frac{1}{\lamD}\:\sqrt{1+(q\lambda_D)^2+i\:\frac{
\omega}{\oD}},\\
\oD&\equiv \frac{D}{\lamDsqr}.
\end{align}
\esub
For the potential we seek a solution of a form similar to
Eq.~(\ref{eq:nu_general}), $\phi\propto \cos(qy)e^{i\omega t}$,
and substituting this together with Eq.~(\ref{eq:nu_general}) into
Eq.~(\ref{eq:Poisson}) yields
\begin{equation}\label{Poisson_phi1}
(\partial_x^2-q^2) \phi = - \frac{Ze}{\epsilon} \mathcal{C}_1
e^{-\kappa x}\cos(qy)e^{i\omega t}.
\end{equation}
Demanding $\phi(\rrr,t)\big|_{x=\infty}=0$ the solution is
 \begin{equation}\label{eq:phi_general}
 \phi=
 \frac{Ze/\epsilon}{q^2-\kappa^2}\Big[\mathcal{C}_1
 e^{-\kappa x} +\mathcal{C}_2 e^{-qx}\Big]\cos(qy)e^{i\omega t},\quad x>0.
 \end{equation}
In the insulating layer we have the following $\cos(qy)e^{i\omega
t}$ modulated general solution to Eq.~(\ref{eq:Laplace}),
\begin{equation}
\phi= \Big[\mathcal{C}_3e^{-q x} +\mathcal{C}_4
e^{qx}\Big]\cos(qy)e^{i\omega t}\:,\: -d<x<0.
\end{equation}
In order to determine $\mathcal{C}_1$, $\mathcal{C}_2$,
$\mathcal{C}_3$, and $\mathcal{C}_4$ we first consider the
boundary condition for the current. Applying the Debye--H\"{u}ckel
approximation to the second term in Eq.~(\ref{boundary_current})
and forming the difference of the "$\pm$" solutions we arrive at
\begin{equation}
0 =  \partial_x \Big[\nu(\rrr,t) + \frac{C_D}{Ze\lambda_D}
\phi(\rrr,t)\Big]\Big|_{x=0}.
\end{equation}
Together with the boundary condition for $\phi$ at $x=-d$ as well
as the continuity of $\phi$ and $\epsilon\partial_x\phi$ at $x=0$
we may determine the constants straightforwardly keeping in mind
that $\kappa$ depends on $q$ and $\omega$. For the coefficients in
the electrolyte we get
\begin{subequations}
\begin{equation}
\mathcal{C}_1= \frac{qV_0}{Ze}\:\frac{\kappa}{q}\:
C_\text{eff}(\omega),
\end{equation}
and
\begin{equation}
\mathcal{C}_2= -i \frac{ \omega}{\oD}\:
\frac{\kappa}{q}\:\mathcal{C}_1,
\end{equation}
while for the insulator the coefficients have a similar, but less
compact form. Above,
 \begin{align}\label{eq:Ceff3}
 C_{\rm
 eff}^{-1}(\omega)&=(\kappa\lambda_D)^2\frac{\sinh(qd)}{qd}\:C_s^{-1}\\
 &\quad +\kappa\lambda_D
 \frac{q\lambda_D(q\lambda_D+\kappa\lambda_D)+i\:\frac{
 \omega}{\oD}}{q\lambda_D(q\lambda_D+\kappa\lambda_D)}\:
 \cosh(qd)\:C_D^{-1}\nonumber,
 \end{align}
 \end{subequations}
which satisfies the definition in Eq.~(\ref{eq:Ceff}) and reduces
to Eq.~(\ref{eq:Ceff2}) in the DC limit. From the general solution
for the potential in the electrolyte, Eq.~(\ref{eq:phi_general}),
we may now in more detail examine the constraints on $V_0$ for the
Debye--H\"{u}ckel approximation to be valid. Straightforward
calculations show that $\max\{\phi\} \ll \kB T/Ze$ corresponds to
$V_0\ll V_T\equiv(1+C_D/C_s) \kB T/Ze$ for $qd\ll 1$,
$q\lambda_D\ll 1$, and low frequencies.

\subsection{Long-period and low-frequency modulation}
\label{Sec:longperiod}

Next, we consider the regime where the spatial period of the
modulation is much longer than all other length scales, \ie,
$q\lambda_D\ll 1$ and $qd\ll 1$. We also assume that
$\omega\ll\oD$ so that $\kappa \simeq 1/\lambda_D$. In this limit
we get
\begin{equation}
\nu= -\frac{ q\sigma_{\scriptscriptstyle \infty} V_0}{
Ze\lambda_D}\frac{1}{\omega^*+i\omega}
e^{-x/\lambda_D}\cos(qy)e^{i\omega t}+{\cal O}([q\lambda_D]^2),
\end{equation}
and
\begin{equation}
\label{eq:phi} \phi= V_0 \frac{i\omega}{\omega^*+i\omega}
e^{-qx}\cos(qy)e^{i\omega t}+{\cal O}(q\lambda_D),
\end{equation}
where we have used the notation of Ajdari \cite{Ajdari:00a}
 \bsub
 \begin{alignat}{2}
 &\textrm{resonance frequency:}& \;\;
 \omega^*&=q\lamD(1+\delta)\:\oD,\label{eq:omegastar}\\
 &\textrm{conductivity:}&
 \sigma_{\scriptscriptstyle \infty}
 &=[\sigma^++\sigma^-]\big|_{\infty}=\epsilon\oD,\\
 &\textrm{capacitance ratio:}& \delta &= \frac{C_D}{C_s}.
 \end{alignat}
\esub
These results are equivalent to those in
Ref.~\onlinecite{Ajdari:00a} if we similarly to
Eq.~(\ref{eq:Ceff}) introduce the Debye layer surface charge
$\sigma_D(y) = Ze\int_0^\infty dx\: \nu(x,y)$.

\subsection{Body-force}

Until this point we have used the exponential notation for the
temporal dependence. However, since the body-force is essentially
non-linear in the electrical potential/density [see last term in
Eq.~(\ref{eq:NS})] we have to take the real part to get the
body-force, \ie,  $\FFF=-Ze\nu \nablabf\phi= -Ze{\rm
Re}\{\nu\}{\rm Re}\{\nablabf\phi\}$ so that we get
\begin{subequations}
\begin{align}
\FFF &=\frac{\eta v^{{}}_1}{\lambda_D^2}\: \frac{\cos(2\omega
t+\varphi)}{\frac{\omega}{\omega^*}
+\frac{\omega^*}{\omega}}\:e^{-x/\lambda_D}\label{eq:bodyforce}\\
&\quad\times \big[2\cos^2(qy)\textbf{e}_x+\sin(2qy)\textbf{e}_y
\big] +{\cal O}\big([q\lambda_D]^2\big)\nonumber
\end{align}
where following Ref.~\onlinecite{Ajdari:00a} we have introduced
\begin{equation}v^{{}}_1 \equiv \frac{q\epsilon
V_0^2}{4\eta(1+\delta)}\label{eq:v1}
\end{equation}
and the frequency dependent phase shift
\begin{equation}
\varphi=-\arctan\left(
\frac{\omega}{2\omega^*}-\frac{\omega^*}{2\omega}\right).
\end{equation}
\end{subequations}
In the derivation of Eq.~(\ref{eq:bodyforce}) we have used that
 \begin{equation}
 {\rm Re}\bigg\{\frac{e^{i\omega t}}{i\omega+\omega^*}\bigg\}{\rm
 Re}\bigg\{\frac{i\omega e^{i\omega t}}{i\omega+\omega^*}\bigg\}
 =\frac{-\cos(2\omega t+\varphi)}{2
 \omega^*\big(\frac{\omega}{\;\omega^*}
 +\frac{\;\omega^*}{\omega}\big)}.
 \end{equation}
At low frequencies, $\FFF\propto \omega$, it becomes maximal at
the resonance frequency $\omega^*$, and then it falls off again at
higher frequencies. We note that $\lim_{\omega\rightarrow
0}\FFF={\cal O}\big([q\lambda_D]^2\big)$, but this small force
will just be balanced by a pressure gradient so that
$\lim_{\omega\rightarrow 0}\vvv=\textbf{0}$ and
$\lim_{\omega\rightarrow 0}\textbf{i}^\pm=\textbf{0}$.

\begin{figure*}
\centerline{\includegraphics[width=\textwidth]{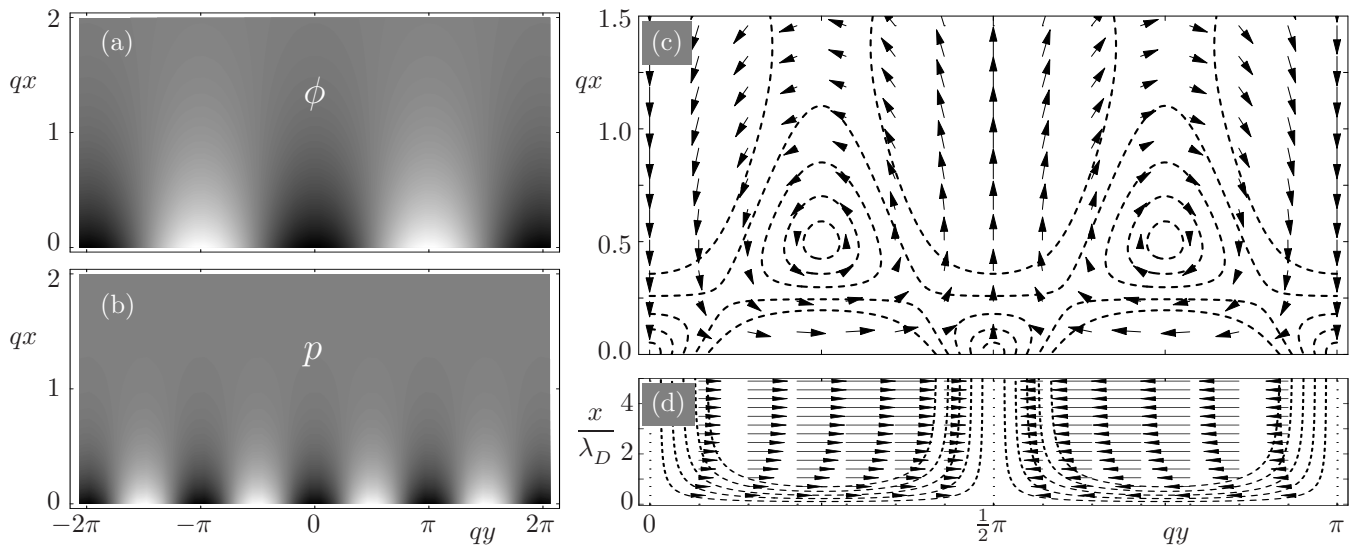}}
\caption{\label{fig:flowpressure} The potential $\phi$, pressure
$p$, and velocity field $\vvv$. Panel (a) shows a gray scale plot
of the amplitude of the potential $\phi$ as a function of $qx$ and
$qy$, Eq.~(\ref{eq:phi}), and panel (b) the pressure $p$,
Eq.~(\ref{eq:pbulk}). Notice the period doubling in the pressure
compared to the electric potential. Panel (c) shows a snapshot of
the harmonically oscillating velocity field $\vvv$ in the bulk,
Eq.~(\ref{eq:vbulk}), and panel (d) likewise in the Debye layer,
Eq.~(\ref{eq:vDebye}). The flow pattern contains rolls, which are
indicated by contours of constant velocity (dashed lines).}
\end{figure*}

\subsection{Linearized flow in quasi-steady state}
\label{Sec:fluidflow}

In order to solve the Navier--Stokes equation, Eq.~(\ref{eq:NS}),
we note that for a body-force of small magnitude and with slow
temporal variation the fluid response is linear and the flow will
approximately be at steady state at each moment in time. We begin
by comparing the inertial terms on the left-hand side (LHS) with
the viscous term (second term) on the right-hand side (RHS). The
body force has a characteristic frequency $\omega$ and two
characteristic length scales $\lambda_D$ and $q^{-1}$ for the $x$
and $y$-directions, respectively. Since $\partial_t$ essentially
gives a factor of $\omega$, and $\nablabf$ essentially gives
$\lambda_D^{-1} \textbf{e}_x+q \textbf{e}_y$, we can show that the
viscous term dominates over the LHS when $\omega\ll \omega_c$
where
\begin{equation}
\omega_c \equiv \frac{\eta}{\rho}\:\min\{q^2,\lambda_D^{-2}\}.
\end{equation}
For $q\lambda_D\ll 1$ this means that $\omega_c =
\frac{\eta}{\rho}\:q^2$. In this way, for small Reynolds numbers,
we get
\begin{equation}
\textbf{0} =  -\nablabf p + \eta \nablabf^2 \vvv+\FFF,\quad
\omega\ll\omega_c \label{eq:NS2}
\end{equation}
which is the resulting quasi-steady flow problem which is linear
in the velocity field. Slip-velocity approaches usually rely on
this equation, see Ref.~\onlinecite{Ajdari:00a} and references
therein. However, Eqs.~(\ref{eq:NS2}) and (\ref{eq:bodyforce}) can
actually be solved exactly with a solution given by

\begin{widetext}
\begin{subequations}
\begin{align}\label{eq:v_unified}
 \vvv(\rrr,t) &= v_1\:
  \frac{\cos(2\omega
  t+\varphi)}{\frac{\omega}{\omega^*}+\frac{\omega^*}{\omega}}\:
 e^{-2 q x}  \Big[
  2q\lambda_D\mathcal{G}_1(\lambda_D^{-1})
   \cos(2qy)\mathbf{e}_x +\mathcal{G}_1(2q)
   \sin(2qy)\mathbf{e}_y
 \Big]\\
 &=
 v_1\:
  \frac{\cos(2\omega t+\varphi)}{\frac{\omega}{\omega^*}
  +\frac{\omega^*}{\omega}}\:  \Big[-2qx e^{-2 q x}
   \cos(2qy)\mathbf{e}_x+\big\{(1-2qx)e^{-2 q x}
   -e^{-x/\lambda_D}\big\}
   \sin(2qy)\mathbf{e}_y
\Big] +{\cal O}(q\lambda_D)  \label{eq:v_unifiedB}
\end{align}
\end{subequations}
\begin{subequations}
\begin{align}
\label{eq:p_unified}
 p(\rrr,t) &= -4q \eta v_1
  \frac{\cos(2\omega t+\varphi)}{\frac{\omega}{\omega^*}
  +\frac{\omega^*}{\omega}}\: e^{-2qx}\Bigg(\frac{1+
  \mathcal{G}_2(2q)}{(1+2q\lambda_D)^2}\cos(2qy) +
  \mathcal{G}_2(0)\Bigg),\\
    &= -4q \eta v_1
  \frac{\cos(2\omega t+\varphi)}{\frac{\omega}{\omega^*}+
  \frac{\omega^*}{\omega}}\: \Bigg(e^{-2qx}\cos(2qy) +
  \frac{e^{-x/\lambda_D}}{4q\lambda_D}\big[1+
  \cos(2qy)\big]
   \Bigg)+{\cal O}(q\lambda_D), \label{eq:p_unifiedB}
\end{align}
\end{subequations}
\end{widetext}
 as may be verified by direct insertion. Above,
\begin{subequations}
\begin{align}
\mathcal{G}_1(k)&=
\frac{1-2q\lambda_D}{\left[1-(2q\lambda_D)^2\right]^2}\nonumber\\
&\quad\times\left(1- e^{-(\lambda_D^{-1}-2q)x}-(1-2q\lambda_D)k
x\right)
\end{align}
and
\begin{equation}
\mathcal{G}_2(k)=\frac{1+k\lambda_D}{4q\lambda_D}\:e^{-(\lambda_D^{-1}-2
q) x}
\end{equation}
\end{subequations}
have been introduced. As seen the flow decays exponentially over a
length scale of $1/q$ when $\omega\ll\omega_c$. When the frequency
becomes comparable to or larger than $\omega_c$ we have competing
length scales since the $\partial_t\vvv$ term introduces an
additional length scale, $(\eta/\rho\omega)^{1/2}$, which as
mentioned becomes $(\omega_c/\omega)^{1/2}\times 1/q$ for
$q\lambda_D\ll 1$. So in the above expressions for $\vvv$ and $p$
we expect that the spatial cut-off length changes from $(2q)^{-1}$
to $\Lambda$ with
\begin{equation}
\Lambda(\omega) \sim \frac{1}{2q}\:
\min\Big\{1,\sqrt{\omega_c/\omega}\:\Big\}.
\end{equation}
Even for $\omega\ll\omega^*$ the condition $\omega\ll\omega_c$ is
not necessarily fulfilled. In fact, for the numbers in
Table~\ref{tab:parameters} we have $\omega_c < \omega^*<\omega_D$
so at resonance $2q\Lambda(\omega^*)=\sqrt{\omega_c/\omega^*}\ll
1$.

\subsection{Flow and separation of length scales}
As mentioned above the flow is typically analyzed by slip-velocity
approaches and here we show how such an approach gives an
asymptotic solution in full agreement with the exact solution. We
study the flow over a $\lambda_D$-scale at the boundary first and
then a $q^{-1}$-scale.
 For this boundary
layer approach, we assume that for $x \lesssim 3\lambda_D$, we
have $v_x \approx 0$. Solving for the pressure and substituting
into the $y$-component of Eq.~(\ref{eq:NS2}) we get
\begin{equation}\label{eq:vDebye}
v_y = v_s(y,t)\: \big(1-e^{-x/\lambda_D}\big)+ {\cal
O}(q\lambda_D),\:x \lesssim 3\lambda_D
\end{equation}
with the prefactor
\begin{equation}\label{eq:vslip}
v_s(y,t) \equiv v^{{}}_1\:\frac{ \cos(2 \, \omega \, t +
\varphi)}{\frac{\omega}{\omega^*}+\frac{\omega^*}{\omega}}\:
\sin(2\,q\,y).
\end{equation}
In the limit $x \gtrsim 3\lambda_D$ and $q\lambda_D\ll 1$ the
velocity $v_s$ can be interpreted as a slip-velocity at the wall
acting as a conveyor belt for the bulk fluid, see
Fig.~\ref{fig:flowpressure}(d).

For $x\gtrsim 3\lambda_D$ we have that $\mathbf{F}$ is
exponentially suppressed and we solve Eq.~(\ref{eq:NS2}) together
with Eq.~(\ref{eq:incompressible}) and the boundary condition
\begin{equation}
\vvv( \rrr,t) \big|_{x=0}=v_s(y,t)\:\textbf{e}_y.
\end{equation}
To lowest order in $q\lambda_D$ this gives
 \begin{align}\label{eq:vbulk}
 \vvv &\simeq v^{{}}_1\: \frac{ \cos(2 \, \omega \, t
 + \varphi)}{\frac{\omega}{\omega^*}+
 \frac{\omega^*}{\omega}}\:e^{-2qx}\\
 &\quad \times \Big(- 2qx \cos(2qy)\textbf{e}_x +
 (1-2qx)\sin(2qy)\textbf{e}_y\Big),\nonumber
 \end{align}
and
 \begin{equation}\label{eq:pbulk}
 p\simeq -4 q \eta  v^{{}}_1 \: \frac{ \cos(2\omega t+
 \varphi)}{\frac{\omega}{\omega^*}+\frac{\omega^*}{\omega}}\:
 e^{-2qx}\cos(2qy).
 \end{equation}
If we now substitute into Eq.~(\ref{eq:NS}) we get
$\textrm{RHS}-\textrm{LHS}\propto e^{-x/\lambda_D}+{\cal
O}(\omega/\oD)+{\cal O}([q\lambda_D]^2)$ which shows that
Eqs.~(\ref{eq:vbulk}) and~(\ref{eq:pbulk}) are indeed excellent
approximations to the full solution of the non-linear
time-dependent Navier--Stokes equation, Eq.~(\ref{eq:NS}), for
$x\gtrsim 3\lambda_D$. For the incompressibility constraint,
Eq.~(\ref{eq:incompressible}), our solution gives $\nablabf \cdot
\vvv={\cal O}([q\lambda_D]^2)$. In Fig.~\ref{fig:flowpressure}(c)
we show a plot of the velocity-field, Eq.~(\ref{eq:vbulk}), along
with the contours for constant velocity.

We note that in the limit $x\gtrsim 3 \lamD$ the exact solutions,
Eqs.~(\ref{eq:v_unifiedB}) and (\ref{eq:p_unifiedB}) reduce to
Eqs.~(\ref{eq:vbulk}) and (\ref{eq:pbulk}) for the bulk.

\section{Full numerical solution}
\label{sec:FullNumerics}

In this section we present results from numerical finite element
simulations (Femlab) of the five coupled equations,
Eqs.~(\ref{eq:Poisson}) to~(\ref{eq:incompressible}), with the
boundary conditions in Eqs.~(\ref{eq:BCphiSurf}) to
(\ref{eq:boundary_density}). For simplicity we assume a low
Reynolds number so that we can neglect the inertial term,
$(\vvv\cdot\nablabf)\vvv$, in Eq.~(\ref{eq:NS}). This provides
full access to the temporal and spatial evolution of the physical
quantities $n^{\pm}$, $\phi$, $\textbf{i}^\pm$, $\bf v$, and $p$.
The spatially periodic problem (in the $y$-direction) would
typically be handled by applying periodic boundary conditions to
the unit-cell (\eg, $0<y<2\pi/q$). However, due to the symmetry of
$V_\textrm{ext}$ the computational domain can be reduced to
$0<y<\pi/q$ with homogeneous Dirichlet or Neumann boundary
conditions. For the $x$ direction our domain is cut off at a
distance $x=6\pi/q$ from the interface using Dirichlet boundary
conditions for the fields. Near the interface to the insulator,
$0<x \lesssim 3\lambda_D$, we employ a structured grid to resolve
the Debye-layer. For the temporal evolution we employ the Femlab
time-stepper directly starting from initial solutions at $t=0$
which are zero everywhere. The duration of the transient depends
on the inertia in the system, but typically the temporal harmonic
state is fully evolved after a time of the order
$10-100\times\omega^{-1}$. For our simulations we have used the
typical values in Table~\ref{tab:parameters} except for the
modulation where we have used $q^{-1}=10^{-6}\,{\rm m}$ and
consequently $\omega^*=\omega_c=10^6\,{\rm rad/s}$. For the
external potential $V_\textrm{ext}$ we have used the imaginary
part of Eq.~(\ref{eq:Vext}) [which is zero at $t=0$] with an
amplitude $0<V^{{}}_0<10$~V and a driving frequency
$10^4\:\textrm{rad/s} < \omega < 10^8\:\textrm{rad/s}$.

\begin{figure}
\centerline{\includegraphics[width=\columnwidth]{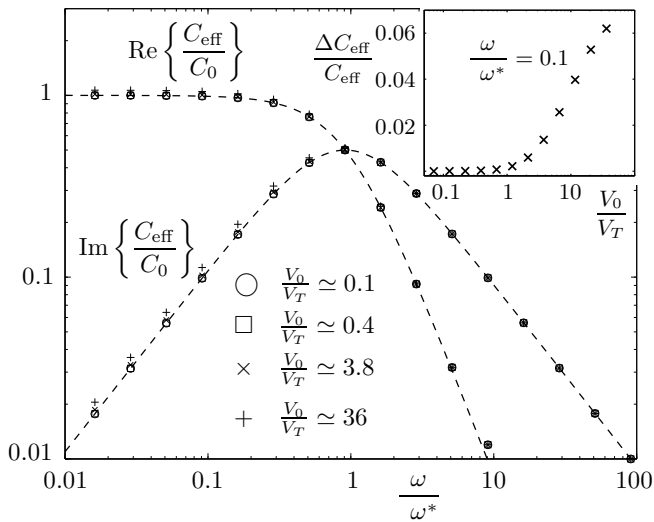}}
\caption{\label{fig:capacitance} The real and the imaginary parts
of the normalized effective capacitance
$C^{{}}_\textrm{eff}(V^{{}}_0,\omega)/C^{{}}_0$ versus the
normalized frequency $\omega/\omega^*$ for four different values
of the voltage amplitude $V^{{}}_0/V^{{}}_T$. The dashed lines
show the analytical results of the linear theory,
Eq.~(\ref{eq:Ceff3}). The inset shows the relative deviation
(below resonance) of the full numerical solution for
$C^{{}}_\textrm{eff}(V^{{}}_0)/C^{{}}_0$ from the
Debye--H\"{u}ckel result as a function of $V^{{}}_0/V_T$.}
\end{figure}

In order to directly compare our numerical results to the
linearized theory we normalize frequencies with the resonance
frequency $\omega^*$, velocities by $v_1$, voltages by the thermal
voltage $V_T$, and capacitances by $C_0$.

Figure~\ref{fig:capacitance} shows numerical results for the
effective capacitance $C_\textrm{eff}$, see definition in
Eq.~(\ref{eq:Ceff}), as a function of the frequency for varying
amplitudes $V_0$ of the external voltage. The dashed line shows
the corresponding analytical result from Eq.~(\ref{eq:Ceff3}). As
seen there is a good agreement between numerical results and our
analytical predictions even for $V_0>V_T$ where the
Debye--H\"{u}ckel approximation is typically expected to work
poorly. We furthermore note that at low frequencies $\textrm{Re}
\{C_\textrm{eff}\}$ approaches $C_0$ in full agreement with the
analytics and the log-log plot also reveals two distinct regimes
for $\omega<\omega^*$ and $\omega>\omega^*$. In fact, dissipation
is maximal exactly at the resonance frequency $\omega^*$ predicted
by the linear theory. The inset shows the relative error of the
Debye--H\"{u}ckel approximation which at large voltages saturates
at a value of the order $(1+\delta)^{-1}$, here equal to 0.09.

\begin{figure*}
\centerline{\includegraphics[width=2\columnwidth]{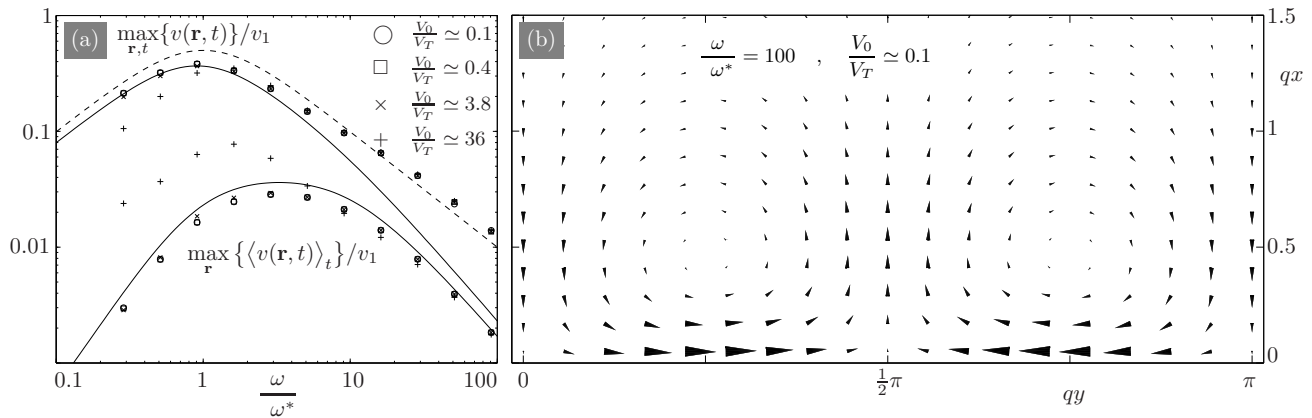}}
\caption{\label{fig:VtimeAvr} Numerical simulations of flow. Panel
(a) shows the maximal velocity of the harmonic flow
$\max_{\rrr,t}\{v(\rrr,t)\}$ and the maximal velocity of the
time-averaged flow $\max_\rrr\big\{\big<v(\rrr,t)\big>_t\big\}$,
both versus normalized frequency. The solid lines show exact
results within the Debye--H\"{u}ckel
approximation~\cite{exactanalytics} and the dashed line shows
$(\frac{\omega}{\omega^*}+
 \frac{\omega^*}{\omega})^{-1}$ as suggested by Eq.~(\ref{eq:vbulk}). Panel (b) shows an example of the time-averaged velocity field $\big<\vvv(\rrr,t)\big>_t$. }
\end{figure*}

The linear theory predicts a harmonic velocity field with a
vanishing time average and our numerical simulations confirm this
low-frequency dynamics, see $\max^{{}}_{\rrr,t}\big\{ v
(\rrr,t)\big\}$ in Fig.~\ref{fig:VtimeAvr}(a). The corresponding
solid line shows exact results within the Debye--H\"{u}ckel
approximation~\cite{exactanalytics} and the dashed line shows
$(\frac{\omega}{\omega^*}+
 \frac{\omega^*}{\omega})^{-1}$ as suggested by
 Eq.~(\ref{eq:vbulk}). As expected the induced harmonic motions peaks at the resonance
frequency $\omega^*$ with a characteristic speed $v_1$,
Eq.~(\ref{eq:v1}). However, in the high-frequency dynamics we
observe the co-existence of a small, but non-vanishing
time-averaged component, $0< \max^{{}}_\rrr \big\{\langle v
(\rrr,t)\rangle^{{}}_t\big\}\ll v_1 $. Fig.~\ref{fig:VtimeAvr}(a)
shows $\max^{{}}_\rrr \big\{\langle v
(\rrr,t)\rangle^{{}}_t\big\}$ as a function of frequency for
different external voltages. The corresponding corresponding solid
line shows exact results within the Debye--H\"{u}ckel
approximation~\cite{exactanalytics}. Panel (b) shows a particular
example of the time-averaged velocity field
$\langle\vvv(\rrr,t)\rangle^{{}}_t$.

\section{Discussion}
\label{sec:discussion}

We have analyzed the full non-equilibrium electro-hydrodynamics of
the Debye screening layer that arise in an aqueous binary solution
near a planar wall when applying a spatially modulated ac-voltage
$V_0\cos(qy)e^{i\omega t}$, Eq.~(\ref{eq:Vext}). Using first order
perturbation theory we have obtained analytic solutions for the
pressure and velocity fields of the electrolyte and for the
electric potential. Our analytical solution applies to the
low-frequency Debye--H\"{u}ckel regime where the amplitude $V_0$
of the external potential is lower than the thermal voltage $V_T$
and the driving frequency $\omega$ is lower than the inverse
response-time of the electrolyte $\oD=\sigma_{\scriptscriptstyle
\infty}/\epsilon$ (see Secs.~\ref{Sec:debye-huckel}
and~\ref{Sec:longperiod}). It should be noted that our analysis
does not cover the special case of suddenly applied step voltages,
where the system selects its own intrinsic time scale different
from the external time scale $1/\omega$~\cite{Bazant:04a}.

Furthermore, we have limited ourselves to the diffusive regime
where convection can be neglected corresponding to a sufficiently
low driving amplitude, $V_0\ll V_c$ where $V_c\equiv
\sqrt{(1+\delta)\eta D/\epsilon}$ is a convective voltage (see
first paragraph of Sec.~\ref{Sec:diffusion}, $\vvv \sim v^{{}}_1
\textbf{e}_y$ in the Debye layer). We have also considered the
low-frequency regime $\omega\ll\omega_c$ where viscosity dominates
over inertia (see Sec.~\ref{Sec:fluidflow}).

Finally, we have considered the limit with the spatial modulation
being much longer than all other length scales in the problem,
\ie, $qd\ll 1$ and $q\lambda_D\ll 1$ (see
Sec.~\ref{Sec:longperiod}). In summary this means that the
analytical studies of the effect of Eq.~(\ref{eq:Vext}) are valid
in the limits
\begin{subequations}
\begin{align}
q&\ll \min\big\{d^{-1},\:\lambda_D^{{-1}}\big\},\\
\omega&\ll \min\big\{\oD,\:\oC\big\},\\
 V_0&\ll
\min\big\{V_T,\: V_c\big\}.
\end{align}
\end{subequations}
As a main result we have supplied a proof for the validity of the
capacitor model. The full dynamics seems however not to be
captured by the capacitor model. Taking the time-average in
Eqs.~(\ref{eq:bodyforce}) and~(\ref{eq:vbulk}) we get
$\big<\FFF\big>_t=\textbf{0}$ and $\big<\vvv\big>_t=\textbf{0}$
(in full agreement with the discussion in
Ref.~\onlinecite{Ajdari:00a}). In contrast, we obtain
$\big<\FFF\big>_t\neq \textbf{0}$ if we begin from
Eqs.~(\ref{eq:nu_general}) and~(\ref{eq:phi_general}) without
expanding in $\omega/\oD$ and $q\lambda_D$, the result being
finite even in the zero-frequency limit. Somewhat similar results
were reported in another non-equilibrium study
\cite{Gonzalez:00a}, though for a different geometry.
Na{\"\i}vely, this observation could suggest that
$\big<\vvv\big>_t\neq \textbf{0}$ contrary to the statement in
Ref.~\onlinecite{Ajdari:00a}. However, by also averaging over the
$y$-direction we get $\big<F_y\big>_{t,y}=0$ suggesting that
$\big<v_y\big>_{t,y}=0$ in agreement with the symmetry arguments
emphasized in Ref.~\onlinecite{Ajdari:00a}. If the finite
$\big<\FFF\big>_t$ does not give the fluid a directional flow
globally, we might speculate that, at high frequencies, it makes
the fluid circulate in non-oscillating vortices, see
Fig.~\ref{fig:VtimeAvr}(a), whereas the fluid is at rest at
zero-frequency -- despite $\big<\FFF\big>_t$ being finite. The
solution to this apparent contradiction lies in the pressure,
which will compensate the body-force at low frequencies. We can
explicitly show that the time-averaged body-force can be written
as
\begin{equation}\label{eq:Ft}
\big<\FFF(\rrr,t)\big>_t=\nablabf p^{{}}_\FFF(\rrr)+{\cal
O}\big([\omega/\oD]^2\big)
\end{equation}
where
\begin{subequations}
 \begin{align}
 p^{{}}_\FFF(\rrr)&\equiv
 \frac{\frac{1}{4}\epsilon q^2 V_0^2}{\left[q\lamD \cosh(qd)+
 \frac{\epsilon}{\epsilon_s}\sqrt{1+
 (q\lamD)^2}\sinh(qd)\right]^2} \nonumber\\
 &\qquad\qquad\times e^{-2 \sqrt{1+(q\lambda_D)^2}
 \:x/\lambda_D}\cos^2(qy),\\
 &=q\eta
 v^{{}}_1\bigg(1+\frac{C_D}{C_s}\bigg)
 \bigg(\frac{C_{\rm eff}(0)}{C_D}\bigg)^2\:\frac{1+
 (q\lambda_D)^2}{(q\lambda_D)^2}\nonumber\\
 &\qquad\quad\times e^{-2 \sqrt{1+(q\lambda_D)^2}
 \:x/\lambda_D}\cos^2(qy).
 \end{align}
 \end{subequations}
The form of Eq.~(\ref{eq:Ft}) suggests that
\begin{equation}\label{eq:v_av}
\big<\vvv(\rrr,t)\big>_t=\textbf{0}+{\cal
O}\big([\omega/\oD]^2\big)
\end{equation}
with $p^{{}}_\FFF$ being a pressure that compensates the
low-frequency part of the body force, see Eq.~(\ref{eq:NS2}). The
time-averaged velocity field $\big<\vvv(\rrr,t)\big>_t$ can be
calculated rigorously and the complex expression (not shown)
agrees fully with Eq.~(\ref{eq:v_av}). At high frequencies we
expect stationary vortices, Fig.~\ref{fig:VtimeAvr}(b), to
co-exist with the harmonic fluid motion illustrated in
Fig.~\ref{fig:flowpressure}(c), whereas at low frequencies the
circulation vanishes and we are left with the pure harmonic
motion. Our time-dependent finite-element simulations in
Fig.~\ref{fig:VtimeAvr} support this picture and similar
time-averaged flow in a slightly different geometry has been
observed both experimentally, theoretically, and numerically
\cite{Green:00a,Gonzalez:00a,Green:02a}.

\section{Conclusion}
\label{sec:conclusion}

Our results provide the theoretical underpinning of the capacitor
model widely used in the
literature~\cite{Ajdari:00a,Brown:00a,Green:00a,Green:02a,Nadal:02a,Morgan:03a,Bazant:04a},
and form a firm starting point for future studies of
electro-kinetic pumps and mixers driven by spatially modulated
surface potentials. In general for large values of $\delta$ we
find that the Debye--H\"{u}ckel approximation works well even at
elevated voltages [agreement within less than $(1+\delta)^{-1}$]
where it is typically expected to work poorly. However, our
non-equilibrium approach has also revealed interesting
short-comings in the capacitor approach for high-frequency
dynamics where static vortices appear along with the harmonic
rolls also predicted by the capacitor model.

\section*{Acknowledgement}

We thank A. Ajdari for stimulating discussions and T.~S. Hansen
for sharing initial numerical results with us. N.~A.~M. is
supported by The Danish Technical Research Council (Grant
No.~26-03-0073) and L.~B. by a Socrates/Erasmus grant from the
European Community.

%\bibliographystyle{apsrev}
%\bibliography{//micdat2/mic/MIFTS/papers/BibTeX/MIFTS01}

\begin{thebibliography}{20}
\expandafter\ifx\csname
natexlab\endcsname\relax\def\natexlab#1{#1}\fi
\expandafter\ifx\csname bibnamefont\endcsname\relax
  \def\bibnamefont#1{#1}\fi
\expandafter\ifx\csname bibfnamefont\endcsname\relax
  \def\bibfnamefont#1{#1}\fi
\expandafter\ifx\csname citenamefont\endcsname\relax
  \def\citenamefont#1{#1}\fi
\expandafter\ifx\csname url\endcsname\relax
  \def\url#1{\texttt{#1}}\fi
\expandafter\ifx\csname
urlprefix\endcsname\relax\def\urlprefix{URL }\fi
\providecommand{\bibinfo}[2]{#2}
\providecommand{\eprint}[2][]{\url{#2}}

\bibitem[{\citenamefont{Yeh et~al.}(1997)\citenamefont{Yeh, Seul, and
  Shraiman}}]{yeh:97a}
\bibinfo{author}{\bibfnamefont{S.~R.} \bibnamefont{Yeh}},
  \bibinfo{author}{\bibfnamefont{M.}~\bibnamefont{Seul}}, \bibnamefont{and}
  \bibinfo{author}{\bibfnamefont{B.~I.} \bibnamefont{Shraiman}},
  \bibinfo{journal}{Nature} \textbf{\bibinfo{volume}{386}}, \bibinfo{pages}{57}
  (\bibinfo{year}{1997}).

\bibitem[{\citenamefont{Ajdari}(2000)}]{Ajdari:00a}
\bibinfo{author}{\bibfnamefont{A.}~\bibnamefont{Ajdari}},
  \bibinfo{journal}{Phys.~Rev.~E} \textbf{\bibinfo{volume}{61}},
  \bibinfo{pages}{R45} (\bibinfo{year}{2000}).

\bibitem[{\citenamefont{Brown et~al.}(2000)\citenamefont{Brown, Smith, and
  Rennie}}]{Brown:00a}
\bibinfo{author}{\bibfnamefont{A.~B.~D.} \bibnamefont{Brown}},
  \bibinfo{author}{\bibfnamefont{C.~G.} \bibnamefont{Smith}}, \bibnamefont{and}
  \bibinfo{author}{\bibfnamefont{A.~R.} \bibnamefont{Rennie}},
  \bibinfo{journal}{Phys.~Rev.~E} \textbf{\bibinfo{volume}{63}},
  \bibinfo{pages}{016305} (\bibinfo{year}{2000}).

\bibitem[{\citenamefont{Green et~al.}(2000)\citenamefont{Green, Ramos,
  Gonz\'{a}lez, Morgan, and Castellanos}}]{Green:00a}
\bibinfo{author}{\bibfnamefont{N.~G.} \bibnamefont{Green}},
  \bibinfo{author}{\bibfnamefont{A.}~\bibnamefont{Ramos}},
  \bibinfo{author}{\bibfnamefont{A.}~\bibnamefont{Gonz\'{a}lez}},
  \bibinfo{author}{\bibfnamefont{H.}~\bibnamefont{Morgan}}, \bibnamefont{and}
  \bibinfo{author}{\bibfnamefont{A.}~\bibnamefont{Castellanos}},
  \bibinfo{journal}{Phys.~Rev.~E} \textbf{\bibinfo{volume}{61}},
  \bibinfo{pages}{4011} (\bibinfo{year}{2000}).

\bibitem[{\citenamefont{Gonz\'{a}lez et~al.}(2000)\citenamefont{Gonz\'{a}lez,
  Ramos, Green, Castellanos, and Morgan}}]{Gonzalez:00a}
\bibinfo{author}{\bibfnamefont{A.}~\bibnamefont{Gonz\'{a}lez}},
  \bibinfo{author}{\bibfnamefont{A.}~\bibnamefont{Ramos}},
  \bibinfo{author}{\bibfnamefont{N.~G.} \bibnamefont{Green}},
  \bibinfo{author}{\bibfnamefont{A.}~\bibnamefont{Castellanos}},
  \bibnamefont{and} \bibinfo{author}{\bibfnamefont{H.}~\bibnamefont{Morgan}},
  \bibinfo{journal}{Phys.~Rev.~E} \textbf{\bibinfo{volume}{61}},
  \bibinfo{pages}{4019} (\bibinfo{year}{2000}).

\bibitem[{\citenamefont{Green et~al.}(2002)\citenamefont{Green, Ramos,
  Gonz\'{a}lez, Morgan, and Castellanos}}]{Green:02a}
\bibinfo{author}{\bibfnamefont{N.~G.} \bibnamefont{Green}},
  \bibinfo{author}{\bibfnamefont{A.}~\bibnamefont{Ramos}},
  \bibinfo{author}{\bibfnamefont{A.}~\bibnamefont{Gonz\'{a}lez}},
  \bibinfo{author}{\bibfnamefont{H.}~\bibnamefont{Morgan}}, \bibnamefont{and}
  \bibinfo{author}{\bibfnamefont{A.}~\bibnamefont{Castellanos}},
  \bibinfo{journal}{Phys.~Rev.~E} \textbf{\bibinfo{volume}{66}},
  \bibinfo{pages}{026305} (\bibinfo{year}{2002}).

\bibitem[{\citenamefont{Nadal et~al.}(2002)\citenamefont{Nadal, Argoul,
  Kestener, Pouligny, Ybert, and Ajdari}}]{Nadal:02a}
\bibinfo{author}{\bibfnamefont{F.}~\bibnamefont{Nadal}},
  \bibinfo{author}{\bibfnamefont{F.}~\bibnamefont{Argoul}},
  \bibinfo{author}{\bibfnamefont{P.}~\bibnamefont{Kestener}},
  \bibinfo{author}{\bibfnamefont{B.}~\bibnamefont{Pouligny}},
  \bibinfo{author}{\bibfnamefont{C.}~\bibnamefont{Ybert}}, \bibnamefont{and}
  \bibinfo{author}{\bibfnamefont{A.}~\bibnamefont{Ajdari}},
  \bibinfo{journal}{Eur.~Phys.~J.~E} \textbf{\bibinfo{volume}{9}},
  \bibinfo{pages}{387} (\bibinfo{year}{2002}).

\bibitem[{\citenamefont{Ajdari}(2002)}]{Ajdari:02a}
\bibinfo{author}{\bibfnamefont{A.}~\bibnamefont{Ajdari}},
  \bibinfo{journal}{Phys.~Rev.~E} \textbf{\bibinfo{volume}{65}},
  \bibinfo{pages}{016301} (\bibinfo{year}{2002}).

\bibitem[{\citenamefont{Gitlin et~al.}(2003)\citenamefont{Gitlin, Stroock,
  Whitesides, and Ajdari}}]{Gitlin:03a}
\bibinfo{author}{\bibfnamefont{I.}~\bibnamefont{Gitlin}},
  \bibinfo{author}{\bibfnamefont{A.~D.} \bibnamefont{Stroock}},
  \bibinfo{author}{\bibfnamefont{G.~M.} \bibnamefont{Whitesides}},
  \bibnamefont{and} \bibinfo{author}{\bibfnamefont{A.}~\bibnamefont{Ajdari}},
  \bibinfo{journal}{Appl.~Phys.~Lett.} \textbf{\bibinfo{volume}{83}},
  \bibinfo{pages}{1486} (\bibinfo{year}{2003}).

\bibitem[{\citenamefont{Studer et~al.}(2004)\citenamefont{Studer, P\'{e}pin,
  Chen, and Ajdari}}]{Studer:04b}
\bibinfo{author}{\bibfnamefont{V.}~\bibnamefont{Studer}},
  \bibinfo{author}{\bibfnamefont{A.}~\bibnamefont{P\'{e}pin}},
  \bibinfo{author}{\bibfnamefont{Y.}~\bibnamefont{Chen}}, \bibnamefont{and}
  \bibinfo{author}{\bibfnamefont{A.}~\bibnamefont{Ajdari}},
  \bibinfo{journal}{Analyst} \textbf{\bibinfo{volume}{129}},
  \bibinfo{pages}{944} (\bibinfo{year}{2004}).

\bibitem[{\citenamefont{Cahill et~al.}(2004)\citenamefont{Cahill, Heyderman,
  Gobrecht, and Stemmer}}]{Cahill:04a}
\bibinfo{author}{\bibfnamefont{B.~P.} \bibnamefont{Cahill}},
  \bibinfo{author}{\bibfnamefont{L.~J.} \bibnamefont{Heyderman}},
  \bibinfo{author}{\bibfnamefont{J.}~\bibnamefont{Gobrecht}}, \bibnamefont{and}
  \bibinfo{author}{\bibfnamefont{A.}~\bibnamefont{Stemmer}},
  \bibinfo{journal}{Phys.~Rev.~E} \textbf{\bibinfo{volume}{70}},
  \bibinfo{pages}{036305} (\bibinfo{year}{2004}).

\bibitem[{\citenamefont{Arnoldus and Hansen}(2004)}]{Arnoldus:04a}
\bibinfo{author}{\bibfnamefont{M.}~\bibnamefont{Arnoldus}} \bibnamefont{and}
  \bibinfo{author}{\bibfnamefont{M.}~\bibnamefont{Hansen}},
  \bibinfo{type}{Bachelor's thesis}, \bibinfo{institution}{Department of Micro
  and Nanotechnology, Technical University of Denmark} (\bibinfo{year}{2004}).

\bibitem[{\citenamefont{Hansen}(2004)}]{Hansen:04a}
\bibinfo{author}{\bibfnamefont{T.~S.} \bibnamefont{Hansen}}, Master's thesis,
  \bibinfo{school}{Department of Chemical Engineering, Technical University of
  Denmark} (\bibinfo{year}{2004}).

\bibitem[{\citenamefont{Ramos et~al.}(1998)\citenamefont{Ramos, Morgan, Green,
  and Castellanos}}]{Ramos:98a}
\bibinfo{author}{\bibfnamefont{A.}~\bibnamefont{Ramos}},
  \bibinfo{author}{\bibfnamefont{H.}~\bibnamefont{Morgan}},
  \bibinfo{author}{\bibfnamefont{N.~G.} \bibnamefont{Green}}, \bibnamefont{and}
  \bibinfo{author}{\bibfnamefont{A.}~\bibnamefont{Castellanos}},
  \bibinfo{journal}{J.~Phys.~D Appl.~Phys.} \textbf{\bibinfo{volume}{31}},
  \bibinfo{pages}{2338} (\bibinfo{year}{1998}).

\bibitem[{\citenamefont{Morgan and Green}(2003)}]{Morgan:03a}
\bibinfo{author}{\bibfnamefont{H.}~\bibnamefont{Morgan}} \bibnamefont{and}
  \bibinfo{author}{\bibfnamefont{N.~G.} \bibnamefont{Green}},
  \emph{\bibinfo{title}{AC Electrokinetics: colloids and nanoparticles}},
  Microtechnologies and Microsystems series (\bibinfo{publisher}{Institute of
  Physics Publishing}, \bibinfo{address}{Bristol}, \bibinfo{year}{2003}).

\bibitem[{\citenamefont{Stone et~al.}(2004)\citenamefont{Stone, Stroock, and
  Ajdari}}]{Stone:04a}
\bibinfo{author}{\bibfnamefont{H.~A.} \bibnamefont{Stone}},
  \bibinfo{author}{\bibfnamefont{A.~D.} \bibnamefont{Stroock}},
  \bibnamefont{and} \bibinfo{author}{\bibfnamefont{A.}~\bibnamefont{Ajdari}},
  \bibinfo{journal}{Annu.~Rev.~Fluid~Mech.} \textbf{\bibinfo{volume}{36}},
  \bibinfo{pages}{381} (\bibinfo{year}{2004}).

\bibitem[{\citenamefont{Bazant and Squires}(2004)}]{Bazant:04b}
\bibinfo{author}{\bibfnamefont{M.~Z.} \bibnamefont{Bazant}} \bibnamefont{and}
  \bibinfo{author}{\bibfnamefont{T.~M.} \bibnamefont{Squires}},
  \bibinfo{journal}{Phys.~Rev.~Lett.} \textbf{\bibinfo{volume}{92}},
  \bibinfo{pages}{066101} (\bibinfo{year}{2004}).

\bibitem[{\citenamefont{Squires and Bazant}(2004)}]{Squires:04a}
\bibinfo{author}{\bibfnamefont{T.~M.} \bibnamefont{Squires}} \bibnamefont{and}
  \bibinfo{author}{\bibfnamefont{M.~Z.} \bibnamefont{Bazant}},
  \bibinfo{journal}{J.~Fluid~Mech.} \textbf{\bibinfo{volume}{509}},
  \bibinfo{pages}{217} (\bibinfo{year}{2004}).

\bibitem[{\citenamefont{Bazant et~al.}(2004)\citenamefont{Bazant, Thornton, and
  Ajdari}}]{Bazant:04a}
\bibinfo{author}{\bibfnamefont{M.~Z.} \bibnamefont{Bazant}},
  \bibinfo{author}{\bibfnamefont{K.}~\bibnamefont{Thornton}}, \bibnamefont{and}
  \bibinfo{author}{\bibfnamefont{A.}~\bibnamefont{Ajdari}},
  \bibinfo{journal}{Phys.~Rev.~E} \textbf{\bibinfo{volume}{70}},
  \bibinfo{pages}{021506} (\bibinfo{year}{2004}).

\bibitem[{exa()}]{exactanalytics}
\bibinfo{note}{The very long analytical expression is not displayed. We assume
  a low P\'{e}clet number and employ the Debye--H\"{u}ckel approximation to
  linearize the electrodynamics and calculate the full expression for $\FFF$
  without expansions in $\omega/\oD$, $q\lambda_D$, and $qd$. Subsequently we
  solve the hydrodynamics in the low Reynolds number limit, but with the
  $\partial_t\vvv$ term taken into account.}

\end{thebibliography}

\end{document}